\documentclass[twocolumn]{aastex631}
\usepackage{amsmath,amssymb,amstext}
\usepackage{gensymb}
\usepackage{multirow}
\usepackage{rotating}
\usepackage{tabularx}
\usepackage{comment}

\usepackage{natbib}

\usepackage{xspace}

\newcommand{\jwst}{\texttt{jwst}\xspace}
\newcommand{\eureka}{\texttt{Eureka!}\xspace}

\newcommand{\jedi}{\texttt{ExoTiC-JEDI}\xspace}

\defcitealias{Lustig-Yaeger2023}{Lustig-Yaeger \& Fu et al. 2023}
\defcitealias{Moran2023}{Moran \& Stevenson et al. 2023}
\defcitealias{May2023}{May \& MacDonald et al. 2023}

\begin{document}

\title{JWST COMPASS: NIRSpec/G395H Transmission Observations of the Super-Earth TOI-836b}

\author[0000-0001-8703-7751]{Lili Alderson} 
\affiliation{School of Physics, University of Bristol, HH Wills Physics Laboratory, Tyndall Avenue, Bristol BS8 1TL, UK}

\author[0000-0003-1240-6844]{Natasha E. Batalha}
\affiliation{NASA Ames Research Center, Moffett Field, CA 94035, USA}

\author[0000-0003-4328-3867]{Hannah R. Wakeford} 
\affiliation{School of Physics, University of Bristol, HH Wills Physics Laboratory, Tyndall Avenue, Bristol BS8 1TL, UK}

\author[0000-0003-0354-0187]{Nicole L. Wallack}
\affiliation{Earth and Planets Laboratory, Carnegie Institution for Science, 5241 Broad Branch Road, NW, Washington, DC 20015, USA}

\author[0000-0002-8949-5956]{Artyom Aguichine}
\affiliation{Department of Astronomy and Astrophysics, University of California, Santa Cruz, CA 95064, USA}

\author[0009-0008-2801-5040]{Johanna Teske} 
\affiliation{Earth and Planets Laboratory, Carnegie Institution for Science, 5241 Broad Branch Road, NW, Washington, DC 20015, USA}

\author[0000-0002-4489-3168]{Jea Adams Redai} 
\affiliation{Center for Astrophysics ${\rm \mid}$ Harvard {\rm \&} Smithsonian, 60 Garden St, Cambridge, MA 02138, USA}

\author[0000-0003-4157-832X]{Munazza K. Alam}
\affiliation{Space Telescope Science Institute, 3700 San Martin Drive, Baltimore, MD 21218, USA}

\author[0000-0002-7030-9519]{Natalie M. Batalha}
\affiliation{Department of Astronomy and Astrophysics, University of California, Santa Cruz, CA 95064, USA}

\author[0000-0002-8518-9601]{Peter Gao} 
\affiliation{Earth and Planets Laboratory, Carnegie Institution for Science, 5241 Broad Branch Road, NW, Washington, DC 20015, USA}

\author[0000-0002-4207-6615]{James Kirk} 
\affiliation{Department ofPhysics, Imperial College London, London, UK}

\author[0000-0003-3204-8183]{Mercedes L\'opez-Morales} 
\affiliation{Center for Astrophysics ${\rm \mid}$ Harvard {\rm \&} Smithsonian, 60 Garden St, Cambridge, MA 02138, USA}

\author[0000-0002-6721-3284]{Sarah E. Moran} 
\affiliation{Department of Planetary Sciences and Lunar and Planetary Laboratory, University of Arizona, Tuscon, AZ, USA}

\author[0000-0003-3623-7280]{Nicholas Scarsdale}
\affiliation{Department of Astronomy and Astrophysics, University of California, Santa Cruz, CA 95064, USA}

\author[0000-0002-0413-3308]{Nicholas F. Wogan}
\affiliation{NASA Ames Research Center, Moffett Field, CA 94035, USA}

\author[0000-0003-2862-6278]{Angie Wolfgang}
\affiliation{Eureka Scientific Inc., 2452 Delmer Street Suite 100, Oakland, CA 94602-3017}

\begin{abstract}

We present two transit observations of the $\sim$870\,K, 1.7\,R$_\oplus$ super-Earth TOI-836b with JWST NIRSpec/G395H, resulting in a 2.8--5.2\,$\mu$m transmission spectrum. 
Using two different reduction pipelines, we obtain a median transit depth precision of 34\,ppm for Visit 1 and 36\,ppm for Visit 2, leading to a combined precision of 25\,ppm in spectroscopic channels 30 pixels wide ($\sim0.02\,\mu$m).
 We find that the transmission spectrum from both visits is well fit by a zero-sloped line by fitting zero-sloped and sloped lines, as well as step functions to our data. Combining both visits, we are able to rule out atmospheres with metallicities $<250\times$Solar for an opaque pressure level of 0.1 bar, corresponding to mean molecular weights of $\lesssim6$\,g\,mol$^{-1}$. We therefore conclude that TOI-836b does not have a H$_2$-dominated atmosphere, in possible contrast with its larger, exterior sibling planet, TOI-836c. We recommend that future proposals to observe small planets exercise caution when requiring specific numbers of transits to rule out physical scenarios, particularly for high metallicities and planets around bright host stars, as \texttt{PandExo} predictions appear to be more optimistic than that suggested by the gains from additional transits implied by our data.

\end{abstract}


%%%%%%%%%%%%%%%%%%%%%%%%%%%%%%%%%%%%%%%%%%%%%%%%%%%%%%%%%%%
\section{Introduction} 
\label{section:intro}
%%%%%%%%%%%%%%%%%%%%%%%%%%%%%%%%%%%%%%%%%%%%%%%%%%%%%%%%%%%

Despite their ubiquity \citep{Batalha2013}, 1--4\,R$_{\oplus}$ exoplanets present some of the most complex challenges for observation and interpretation, hampered further by the lack of solar system counterparts. In this radius regime, exoplanets are typically split between two categories: the larger sub-Neptunes ($>1.8$\,R$_{\oplus}$), which may have hydrogen-rich envelopes \citep{Lopez2013, Buchhave2014}, and the smaller super-Earths, with likely more tenuous (if any) atmospheres \citep{Rogers2015, Rogers2021}, while a dearth of planets exists in between (known as the radius valley, e.g., \citealt{Fulton2017}). Currently, one of the leading theories proposed to explain the radius valley is that the smaller planets were not massive enough to retain their primordial atmospheres \citep[e.g.,][]{Lopez2012}. Both photoevaporation \citep[e.g.,][]{Owen2012, Owen2013} and core-powered mass loss \citep[e.g.,][]{Ginzburg2018} can generate heat-driven hydrodynamic outflows from the upper atmosphere, but it is not clear which mechanism is dominant \citep[e.g.,][]{Rogers2021,Owen2023}. 

Regardless of which category an individual exoplanet may lie in, the densities of planets in this parameter space typically allow for many possible interior compositions \citep[e.g.,][]{Rogers2010, Zeng2019}. Without the ability to directly probe the interiors of these exoplanets, observing their atmospheres remains the only way to understand this population in detail. Indeed, studies suggest that determining atmospheric mean molecular weights or metallicities can help break some of the degeneracies presented by interior structure modelling and constrain bulk compositions \citep{Figueira2009, Rogers2011, Fortney2013}. 

The advent of JWST and its exquisite precision across a wide wavelength range \citep{Ahrer2023, Alderson2023, Feinstein2023, Rustamkulov2023} unlocks the ability to explore small exoplanets in detail.
Broad coverage of the infra-red (IR) enables the detection of a variety of molecular species that were previously inaccessible \citep{Batalha2017}, and critically offers the opportunity to explore wavelengths that are less prone to muting from clouds and hazes that have plagued the observation of small exoplanet atmospheres with the Hubble Space Telescope and ground-based instruments \citep[e.g.,][]{Crossfield2013,Kreidberg2014_gj1214, Louden2017, Kirk2018, Ahrer2023_lrgbeasts}. Of particular interest to the study of super-Earth atmospheres is JWST's NIRSpec/G395H mode \citep[][]{Jakobsen2022,Birkmann2022}. Spanning 2.87--5.14\,$\mu$m at R$\sim$2700, the G395H grating covers spectral features from the major absorption bands of CO$_2$, CO and CH$_4$ as well as a partial band of H$_2$O -- molecules expected to sculpt super-Earth transmission spectra across a variety of metallicities \citep[e.g.,][]{Wordsworth2022}. Furthermore, since super-Earths have predominately been found around bright stars, G395H's brightness limit affords the ability to observe these atmospheres without saturating.  
Assessing the presence or absence of these four molecules provides a zeroth-order assessment of the carbon-to-oxygen ratio (C/O) of the atmosphere \citep{Batalha2023}.

Historically, Hubble and ground-based observations of super-Earth (R$<$1.8\,R$_\oplus$) atmospheres have typically yielded featureless transmission spectra \citep[e.g.,][]{Diamond-Lowe2020, Libby-Roberts2022, Diamond-Lowe2023}, but even with the power of JWST, exploring these exoplanets has not been without challenges. NIRSpec/G395H observations of GJ\,486b show evidence of a deviation from a flat line consistent with either a water-rich atmosphere or with contamination from unocculted starspots \citepalias{Moran2023}. The transmission spectra also showed a consistent offset between the two NIRSpec detectors, potentially due to the superbias subtraction step in the data reduction. Similar conflicting inferences have been seen for GJ\,1132b, where the transmission spectrum obtained during one visit is consistent with a water-dominated atmosphere, while the second visit presents a featureless spectrum. In this case, \citetalias{May2023} find that these discrepancies are most likely due to an unlucky random noise draw.

In order to draw conclusions about the broader small exoplanet population as a whole, the JWST COMPASS (Compositions of Mini-Planet Atmospheres for Statistical Study) Program (GO-2512, PIs N. E. Batalha \& J. Teske) is focusing on observing a statistically motivated sample of 1--3\,R$_{\oplus}$ planets. The program will obtain NIRSpec/G395H transmission spectra of eleven exoplanets, while the full statistical sample includes 12 planets, with four pairs in the same systems.\footnote{Our full sample includes L~98-59c and L~98-59d, the latter of which is being observed by GTO-1224, PI S. Birkmann.} 
The targets were selected from a subset of the $\leq 3\,$R$_{\oplus}$ planets observed as part of the Magellan-TESS Survey \citep[MTS,][]{Teske2021}, in order to understand to what extent small planets have detectable atmospheres, and explore the compositional diversity of the population as a whole. 
Similarly to the MTS targets, the COMPASS targets were chosen using a quantitatively selected sample using a merit function based on R$_\mathrm{P}$, insolation flux, stellar effective temperature, and exposure time required to obtain 30\,ppm precision in an R$\sim$100 NIRSpec/G395H bin at 4\,$\mu$m \citep[see][]{Batalha2023}. Specifically, \citet{Batalha2023} showed that a quantitatively chosen sample was shown to be a useful method for enabling constraints on population-level parameters, which is the ultimate goal of the COMPASS Program. By observing multi-planet systems, the COMPASS Program also has the ability to test a variety of formation and evolution theories that are heavily dependent on insolation flux.

The first multi-planet system observed in the COMPASS Program is that of TOI-836 (HIP 73427), in which two planets are orbiting a bright (J mag $\sim$ 7.58) K-dwarf \citep{Hawthorn2023}. The larger of the planets, the sub-Neptune TOI-836.01 (planet c), has a bulk density consistent with a gaseous envelope, with a radius of 2.59$\pm$0.09~R$_{\earth}$ and a mass of 9.6$\pm$2.7\,M$_{\earth}$, and orbits on a period of 8.59\,days. The smaller TOI-836.02 (planet b) is interior, on a period of 3.81\,days and T$_\mathrm{eq}\sim$870\,K. With a radius of 1.70$\pm$0.06\,R$_{\earth}$ and mass of 4.5$\pm$0.9\,M$_{\earth}$, TOI-836b is a super-Earth at the lower edge of the radius valley, and likely has a much smaller gas fraction. Given their positions respective to the radius valley, the TOI-836 system presents an excellent opportunity to directly compare and contrast the atmospheres of differently-sized exoplanets that formed within the same stellar environment. 
Here, we focus on TOI-836b, presenting the 3--5\,$\mu$m transmission spectrum before taking a broader look at the system as a whole, including the observations of TOI-836c presented in \citet{Wallack2024_836.01}.

In \S \ref{section:obs}, we describe our observations and detail our reduction procedures in \S \ref{section:data_reduction}. We present the transmission spectrum of TOI-836b in \S \ref{section:results}, and interpret the transmission spectrum using 1D radiative-convective atmospheric models in \S \ref{section:interpretation}. Finally, we discuss the implications of our results on the TOI-836 system and for future observations in \S \ref{section:discussion} and summarise our conclusions in \S \ref{section:conclusions}.

%%%%%%%%%%%%%%%%%%%%%%%%%%%%%%%%%%%%%%%%%%%%%%%%%%%%%%%%%%%
\section{Observations} 
\label{section:obs}
%%%%%%%%%%%%%%%%%%%%%%%%%%%%%%%%%%%%%%%%%%%%%%%%%%%%%%%%%%%

We observed two transits of TOI-836b with JWST NIRSpec using the high-resolution (R$\sim$2700) G395H mode, which commenced on March 4 2023 at 18:09 UTC and March 8 2023 at 13:45 UTC, respectively. These visits were coincidentally separated by one orbital period, considerably less than the 22-day rotation period of TOI-836 \citep{Hawthorn2023}. NIRSpec/G395H provides spectroscopy between 2.87--5.14\,$\mu$m across the NRS1 and NRS2 detectors (with a gap between 3.72--3.82\,$\mu$m). Both observations were taken in NIRSpec Bright Object Time Series (BOTS) mode using the SUB2048 subarray, F290LP filter, S1600A1 slit, and NRSRAPID readout pattern.  Each 5.3-hour observation consisted of 5259 integrations with 3 groups per integration, and was designed to cover the 1.8-hour transit and sufficient pre- and post-transit baseline.

\begin{deluxetable}{ll}

\tablewidth{0pt}
\tablehead{\colhead{Property}&\colhead{Value}}
\startdata
K (mag)  & 6.804 $\pm$ 0.018 \\
R$_*$ (R$_\odot$)& 0.665$\pm$0.010\\
T$_*$ (K)& 4552$\pm$154\\
log(g) & 4.743$\pm$0.105\\
$[$Fe/H$]_{*}$& -0.284$\pm$-0.067\\
\hline
Period (days)& 3.81673 $\pm$ 0.00001\\
M$_\mathrm{P}$ (M$_{\earth}$)&  4.5$^{+0.92}_{-0.86}$\\
R$_\mathrm{P}$ (R$_{\earth}$)& 1.704 $\pm$  0.067\\
T$_\mathrm{eq}$ (K)& 871$\pm$36\\
$e$& 0.053$\pm$0.042\\
$\omega$ (\textdegree)&9 $\pm$ 92\\
Semi-major axis (AU) & 0.04220 $\pm$ 0.00093 
\tablecaption{System Properties for TOI-836b. Values used in the light curve fitting are shown in Table \ref{table:fit_table}.}
\enddata 
\label{table:system}
\tablenotetext{}{All values from \cite{Hawthorn2023}}
\end{deluxetable}

%%%%%%%%%%%%%%%%%%%%%%%%%%%%%%%%%%%%%%%%%%%%%%%%%%%%%%%%%%%
\section{Data Reduction} 
\label{section:data_reduction}
%%%%%%%%%%%%%%%%%%%%%%%%%%%%%%%%%%%%%%%%%%%%%%%%%%%%%%%%%%%

To check for consistency and ensure robust conclusions, we reduced the data using two independent pipelines: \texttt{ExoTiC-JEDI} \citep{Alderson2022, Alderson2023} and \texttt{Eureka!} \citep{Bell2022}. Each reduction process is described in detail below and follows similar procedures to other NIRSpec/G395H transmission spectra analyses.

\subsection{ExoTiC-JEDI} \label{section:jedi}
The Exoplanet Timeseries Characterisation - JWST Extraction and Diagnostic Investigator (\jedi) package\footnote{https://github.com/Exo-TiC/ExoTiC-JEDI} performs end-to-end extraction, reduction, and analysis of JWST time-series data from \texttt{uncal} files through to light curve fitting to produce planetary spectra. Throughout, NRS1 and NRS2 data are reduced independently, and each visit is treated separately. In all cases, we tried a variety of values for each reduction parameter, and determined the value which resulted in the smallest out-of-transit scatter in the resulting white light curve.

We begin with a modified version of Stage 1 of the \jwst pipeline \citep[v.1.8.6, context map 1078;][]{Bushouse2022}, performing linearity, dark current and saturation corrections, and using a jump detection threshold of 15. We next perform a custom destriping routine to remove group level 1/$f$ noise, masking the spectral trace 15$\sigma$ from the dispersion axis for each integration, subtracting the median pixel value of non-masked pixels from each detector column in each group. We also perform a custom bias subtraction, building a pseudo-bias image by computing the median of each detector pixel in the first group across every integration in the time series. This median image is then used in place of a bias and subtracted from every group, and was found to improve the out-of-transit scatter for both detector white light curves for both visits \citep[see also][]{Alderson2023}. We then proceed with the standard ramp fitting step. \jedi also utilises Stage 2 of the \jwst pipeline to produce the 2D wavelength map needed to obtain the wavelength solution.

In Stage 3 of \jedi, we extract our 1D stellar spectra, performing additional cleaning steps and 1/$f$ correction. Using the standard data quality flags produced by the \jwst pipeline, replacing any pixels flagged as do not use, saturated, dead, hot, low quantum efficiency or no gain value with the median of the neighbouring 4 pixels in each row. To replace any spurious pixels that have not yet been corrected (such as cosmic rays), we identify any that are outliers from their nearest neighbours on the detector, or throughout the time series. We use a 20$\sigma$ threshold in time and a 6$\sigma$ threshold spatially, replacing the problem pixel with the median of that pixel in the surrounding 10 integrations or 20 pixels in the row, respectively. Any remaining 1/$f$ noise and background are removed by subtracting the median unilluminated pixel value from each column in each integration. To extract the 1D stellar spectra, we fit a Gaussian to each column to obtain the centre and width of the spectral trace across the detector, fitting a fourth-order polynomial to each. The spectral trace centres and widths are then smoothed with a median filter of window size 11 to determine the aperture region. For both visits and both detectors, we used an aperture five times the FWHM of the trace, approximately 8 pixels wide from edge to edge. An intrapixel extraction is used to obtain the 1D stellar spectra, where intrapixel is defined as the fraction of the FWHM which falls on each pixel, such that at the edge of the aperture, the flux included from any intersected pixel is equal to the fraction of the pixel within the aperture, multiplied by the total flux value of that pixel. The 1D stellar spectra are then cross-correlated to obtain the $x$- and $y$-positional shifts throughout the observation for use in systematic light curve detrending.

Finally, we fit white light curves for both NRS1 and NRS2, as well as spectroscopic light curves across the full NIRSpec/G395H wavelength range at a variety of resolutions for both visits. For the white light curves (spanning 2.814--3.717\,$\mu$m for NRS1 and 3.824--5.111\,$\mu$m for NRS2), we fit for the system inclination, $i$, ratio of semi-major axis to stellar radius, $a/R_*$, centre of transit time, T$_0$, and the ratio of planet to stellar radii, $R_p/R_*$, holding the period and eccentricity, $e$, and argument of periastron, $\omega$, fixed to values presented in \citet{Hawthorn2023}. The stellar limb darkening coefficients are held fixed to values calculated using the \texttt{ExoTiC-LD} package \citep{Grant2022} based on the stellar T$_{*}$, log(g), and [Fe/H]$_{*}$ presented in \citet{Hawthorn2023} (see Table \ref{table:system}), with Set One of the MPS-ATLAS stellar models \citep{Kostogryz2022, Kostogryz2023} and the non-linear limb darkening law \citep{Claret2000}. We used a least-squares optimiser to fit for a \texttt{batman} \citep{Kreidberg2015} transit model simultaneously with our systematic model $S(\lambda)$, which took the form
$$ S(\lambda) = s_0 + (s_1 \times t) + (s_2 \times x_{s}|y_{s}|) \mathrm{,}$$
where $x_s$ is the $x$-positional shift of the spectral trace, $|y_s|$ is the absolute magnitude of the $y$-positional shift of the spectral trace, $t$ is the time and $s_0, s_1, s_2$ are coefficient terms, as previously used for \jedi analysis in \citetalias{May2023}. For the spectroscopic light curves, we fit for $R_p/R_*$, holding T$_0$, $i$ and $a/R_*$ fixed to the respective white light curve fit value, as shown in Table \ref{table:fit_table}. 

\begin{figure*}
\begin{centering}
\includegraphics[width=\textwidth]{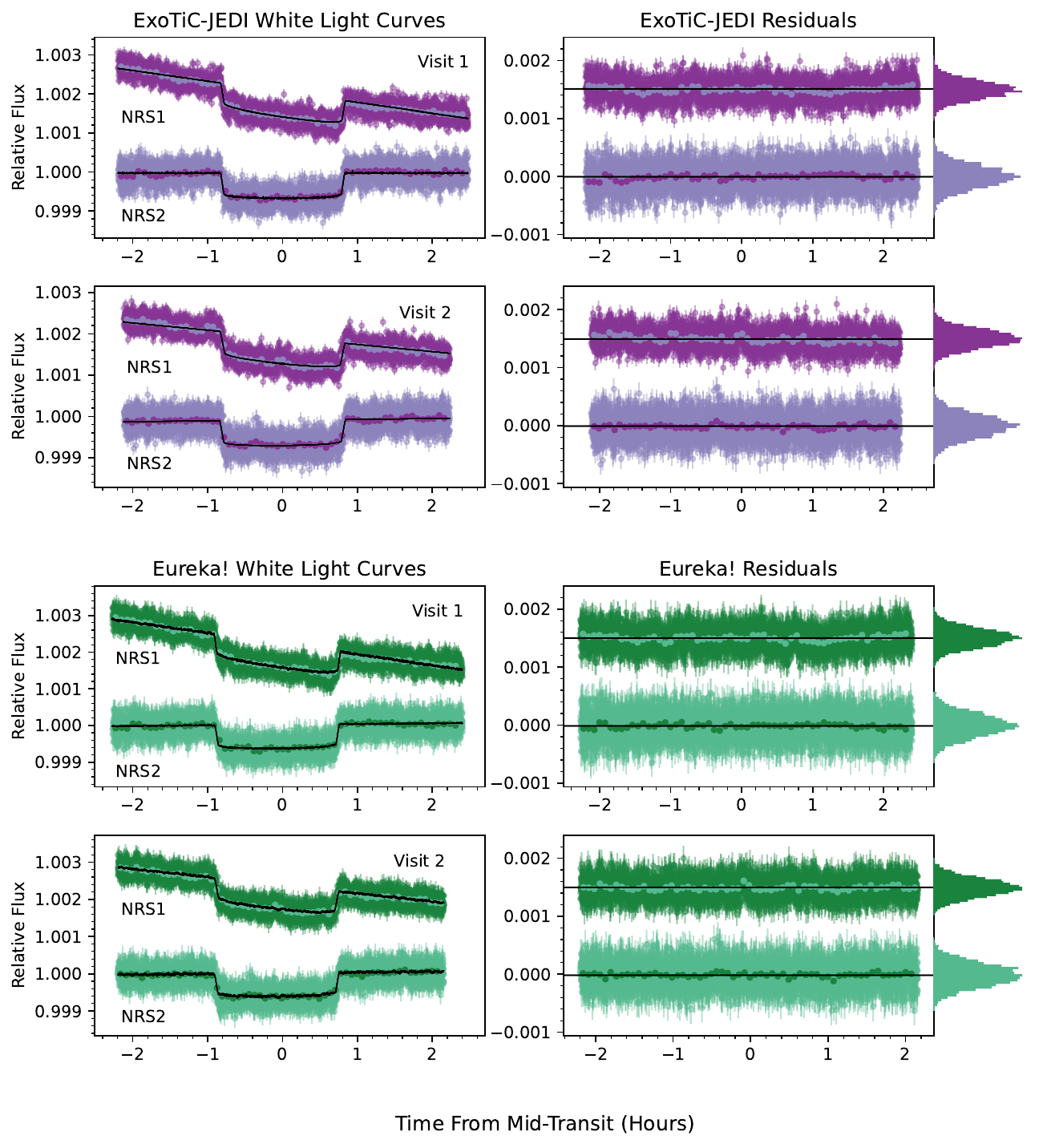}
\caption{White light curves for each reduction, the best-fit models to those white light curves, and the associated residuals for \jedi and \eureka for each visit and detector. Binned light curves and residuals are also shown in the alternate colours. Histograms of the residuals are shown in the rightmost column.} 
\label{figure:wlc}   
\end{centering}
\end{figure*}

For both the white and spectroscopic light curves, we removed any data points that were greater than 4$\sigma$ outliers in the residuals, and refit the light curves until no such points remained. We also rescaled the flux time series errors using the beta value \citep{Pont2006} as measured from the white and red noise values calculated using the \texttt{extra\_functions.noise\_calculator()} in \jedi to account for any remaining red noise in the data. We removed the first 370 integrations ($\sim$ 22 minutes) of visit 1 and the first 440 integrations ($\sim$ 26 minutes) of visit 2 to remove settling ramps at the start of the observations. We additionally removed 259 integrations from the end of visit 1 ($\sim 15$ minutes), and 508 integrations from the end of visit 2 ($\sim 30$ minutes), which removed a slight linear slope in the residuals of NRS1 fits and removed an $\sim$10\,ppm offset between the transit depths of NRS1 and NRS2. The \jedi fitted white light curves and residuals for each visit are shown in Figure \ref{figure:wlc}.

\begin{table*}[]
\centering
\caption{Best fit values for the four individual white light curve fits for \jedi and \eureka as shown in Figure \ref{figure:wlc} and the \eureka joint fit.}
\label{table:fit_table}
\begin{tabular}{ccc|c|c|c|c}
\multicolumn{3}{c|}{}                     & T$_{0}$ (MJD) & $a/R_*$ & $i$ ($\degree$) & R$_p$/R$_*$ \\ \hline
\multicolumn{3}{c|}{\citet{Hawthorn2023} Value}           & 581599.9953$\pm$2e-3  &  --  & 87.57$\pm$0.44 & 0.0235 $\pm$ 0.0013  \\ \hline
%%%%%%%%%%%%%%%%
\multicolumn{1}{c|}{\multirow{4}{*}{\jedi}}   & \multicolumn{1}{c|}{\multirow{2}{*}{Visit 1}} & NRS1 & $60007.86562 \pm7$e${-5}$  & $15.42\pm1.01$   & $88.10\pm0.49$ & 0.02458 $\pm$ 0.00016  \\
\multicolumn{1}{c|}{}                                        & \multicolumn{1}{c|}{}                         & NRS2 & $60007.86552\pm8$e${-5}$   & $15.65\pm1.20$   & $88.19\pm0.58$ & 0.02489 $\pm$ 0.00018  \\
\multicolumn{1}{c|}{}                                        & \multicolumn{1}{c|}{\multirow{2}{*}{Visit 2}} & NRS1 & $60011.68218 \pm 7$e${-5}$ & $14.12\pm1.17$   & $87.45\pm0.59$ & 0.02478 $\pm$ 0.00015  \\
\multicolumn{1}{c|}{}                                        & \multicolumn{1}{c|}{}                         & NRS2 & $60011.68213\pm9$e${-5}$   & $15.01\pm1.38$   & $87.91\pm0.78$ & 0.02448 $\pm$ 0.00017  \\ \hline
%%%%%%%%%%%%%%%%
\multicolumn{1}{c|}{\multirow{8}{*}{\eureka}} & \multicolumn{1}{c|}{\multirow{2}{*}{Visit 1}} & NRS1 & $60007.86570 \pm 6$e${-5}$         & $17.50 \pm 0.77$ & $89.70 \pm 0.82$         & 0.02451 $\pm$ 0.00013 \\
\multicolumn{1}{c|}{}                                        & \multicolumn{1}{c|}{}                         & NRS2 & $60007.86560 \pm 8$e${-5}$           & $15.75 \pm 1.25$   & $88.24 \pm 0.74$ & 0.02489 $\pm$ 0.00018 \\
\multicolumn{1}{c|}{}                                        & \multicolumn{1}{c|}{\multirow{2}{*}{Visit 2}} & NRS1 & $60011.68205 \pm 7$e${-5}$           & $14.12 \pm 1.18$ & $87.45 \pm 0.60$ & 0.02503 $\pm$ 0.00020 \\
\multicolumn{1}{c|}{}                                        & \multicolumn{1}{c|}{}                         & NRS2 & $60011.68207 \pm 9$e${-5}$           & $15.08 \pm1.49$   & $87.94 \pm 1.06$ & 0.02450 $\pm$ 0.00019 \\
%%%%%%%%%%%%%%%%
\multicolumn{1}{c|}{}                                        & \multicolumn{1}{c|}{\multirow{4}{*}{Joint}}   & \multirow{2}{*}{NRS1} & $60007.86552 \pm 5$e${-5}$                     & \multirow{2}{*}{$14.99 \pm 0.84$} & \multirow{2}{*}{$87.89 \pm 0.42$} & \multirow{2}{*}{0.02482 $\pm$ 0.00013} \\
\multicolumn{1}{c|}{}                                        & \multicolumn{1}{c|}{}                         &                       & $60011.68225 \pm 5$e${-5}$                    &                       &                    &     \\
\multicolumn{1}{c|}{}                                        & \multicolumn{1}{c|}{}                         & \multirow{2}{*}{NRS2} & $60007.86548 \pm 6$e${-5}$                   & \multirow{2}{*}{$15.05 \pm0.89$} & \multirow{2}{*}{$87.92 \pm 0.44$} & \multirow{2}{*}{0.02469 $\pm$ 0.00014} \\
\multicolumn{1}{c|}{}                                        & \multicolumn{1}{c|}{}                         &                       & $60011.68221 \pm 6$e${-5}$                   &                        &                       
\end{tabular}
\end{table*}

\subsection{Eureka!}\label{sec:eureka}

For our second independent reduction, we utilise \eureka, an end-to-end pipeline for analysing JWST transiting planet data. We used the default procedures for Stage 1 and Stage 2 from the \eureka wrapper of the \jwst pipeline, following the same procedures as \jedi but using the standard superbias subtraction. We also used the aforementioned \jedi group-level background subtraction to account for 1/$f$ noise. Following this, we use \eureka Stage 3 to extract the stellar spectra and produce the broadband and spectroscopic light curves. Note that while \jedi maintains the slightly curved shape of the NIRSpec/G395H spectral trace throughout spectral extraction, the \eureka pipeline flattens this trace by bringing the centre of mass of each column to the centre of the subarray, allowing for a straight box extraction to be used. \eureka allows for the customisation of a variety of reduction parameters in Stage 3, including the trace extraction width, the region and method for the background extraction, and trace extraction parameters. To find the best combination of values, we tested extraction apertures consisting of combinations of 4-8 pixels from the centre of the flattened trace, background apertures of 8-11 pixels, sigma thresholds for optimal extraction outlier rejection of 10 and 60 (which approximates standard extraction), and two different methods of background subtraction (an additional column-by-column mean subtraction and a full frame median subtraction). We select the final reduction parameters as the combination that minimises the scatter in the resulting white light curves, doing this separately for each detector and each visit. We find that for both visits, both NRS1 and NRS2 favoured an additional column-by-column background subtraction using a sigma threshold of 60 (which approximates a standard box extraction). The optimal aperture half-widths for the trace extraction for NRS1 and NRS2 was 6 pixels for visit 1 and 4 pixels for visit 2. The background subtraction region spans from the upper and lower edge of the detector subarray to an inner bound defined by a number of pixels away from the flattened trace. This inner bound was found to be 8 pixels in both detectors for Visit 1 and 9 pixels for NRS1 and 8 pixels for NRS2 for Visit 2. We then extract white light and 30-pixel binned ($\sim 0.02\,\mu$m, R$\sim$200) spectroscopic light curves for both visits for both detectors.

During the light curve fitting stage, we move away from the \eureka pipeline and utilise a custom light curve fitting code, but refer to this reduction as the ``\eureka'' reduction for simplicity.
We fit each white and spectroscopic light curve separately using \texttt{emcee} \citep{Foreman-Mackey2013}, fitting for $i$, $a/R_*$, T$_0$, and $R_p/R_*$ and fixing the other orbital parameters to those from Table~\ref{table:system} using the \texttt{batman} package \citep{Kreidberg2015}. We utilise quadratic limb-darkening coefficients calculated using \texttt{ExoTiC-LD} and the stellar parameters from Table~\ref{table:system}. We fit our transit model and a systematic model simultaneously, which took the form
$$ S(\lambda) = s_0 + s_1 \times T + s_2 \times X+ s_3 \times Y \mathrm{,}$$
where X and Y are the normalised positions of the trace on the detector and $s_i$ are the free parameters in our systematic noise model. We use an iterative rolling median outlier rejection with a 50 data point-wide window three times on both the white and spectroscopic light curves, removing outliers more than 3$\sigma$ from the median. We initialise our MCMC walkers using the best fit results from a Levenberg–Marquardt least-squares minimisation. We utilise three times the number of free parameters as the number of walkers (resulting in 27 walkers) and run a burn-in of 50,000 steps which is discarded followed by a production run of 50,000 steps, with uninformed priors on all of the parameters. We trim the initial 444 points ($\sim$20 minutes) from all the light curves to remove any initial ramp that may be present, and removed the last 259 points of Visit 1 and 508 points of Visit 2 (see Section \ref{section:jedi}). The fitted white light curves and residuals resulting from the \eureka reduction for each visit are shown in Figure \ref{figure:wlc}, while the fitted white light curve parameters are shown in Table \ref{table:fit_table}. 

During our spectroscopic light curve fits, we once again fit for $i$, $a/R_*$, T$_0$, and $R_p/R_*$, which results in parameters that are consistent with the fitted white light curve values to within 2$\sigma$. We utilise a prior when fitting the $i$, $a/R_*$, T$_0$ for the spectroscopic light curves. To obtain priors for these fits, we utilise the posterior distribution for each free parameter that resulted from the MCMC chains of the white light curve fits for each visit. We use the median 3$\sigma$ values from combining the chains from NRS1 and NRS2 as Gaussian priors for each astrophysical parameter, meaning that the prior represents the combined constraints from NRS1 and NRS2. For $R_p/R_*$, we use a flat uninformed prior to not bias our transmission spectrum.

\subsubsection{Joint Fit of the Eureka! Light Curves}

In order to evaluate the power of combining multiple visits, we also produce a joint fit of the \eureka reduction light curves. Here, we fit both visits simultaneously but continue to separate NRS1 and NRS2 to account for any offsets between the two detectors and the differing systematic effects. We follow the same procedures as for the individual fits (\S \ref{sec:eureka}), but now obtain a universal value for $i$, $a/R_*$, T$_0$, and $R_p/R_*$ for each detector in both the white and spectroscopic light curves.
In the case of T$_0$, we assumed a centre of transit time for each visit and fit for a common offset from this value which applies to both visits for each detector (note that the visits are separated by a single orbital period, see Table \ref{table:fit_table}). 
We again use an MCMC fit, initialising our walkers using a Levenberg-Marquardt least-squares minimizer, with three times the number of walkers as free parameters in our fit (resulting in 42 walkers). For the MCMC we discard a burn-in of 50,000 steps before utilising a production run of 50,000 steps. Our best-fit joint white light curve parameters are shown in Table \ref{table:fit_table} in comparison to those from the individual fits. For the spectroscopic light curves, we once again fit a Gaussian to the posterior distribution for each free parameter which resulted from the MCMC chains, and apply this as our prior for the respective parameters. As with the individual fits, for $R_p/R_*$ we use a flat uninformed prior to not bias our transmission spectrum. For this joint fit, the fitted spectroscopic light curve parameters agree with the white light curve values to within 1$\sigma$ in all wavelength channels.

%%%%%%%%%%%%%%%%%%%%%%%%%%%%%%%%%%%%%%%%%%%%%%%%%%%%%%%%%%%

\section{Transmission Spectrum} 
\label{section:results}

\begin{figure*}
\begin{centering}
\includegraphics[width=\textwidth]{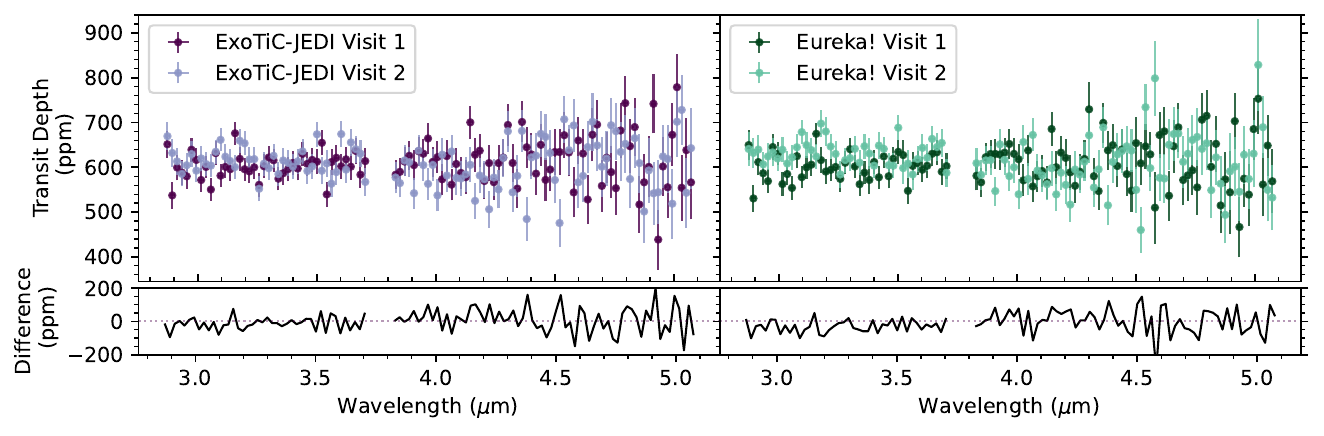}
\includegraphics[width=\textwidth]{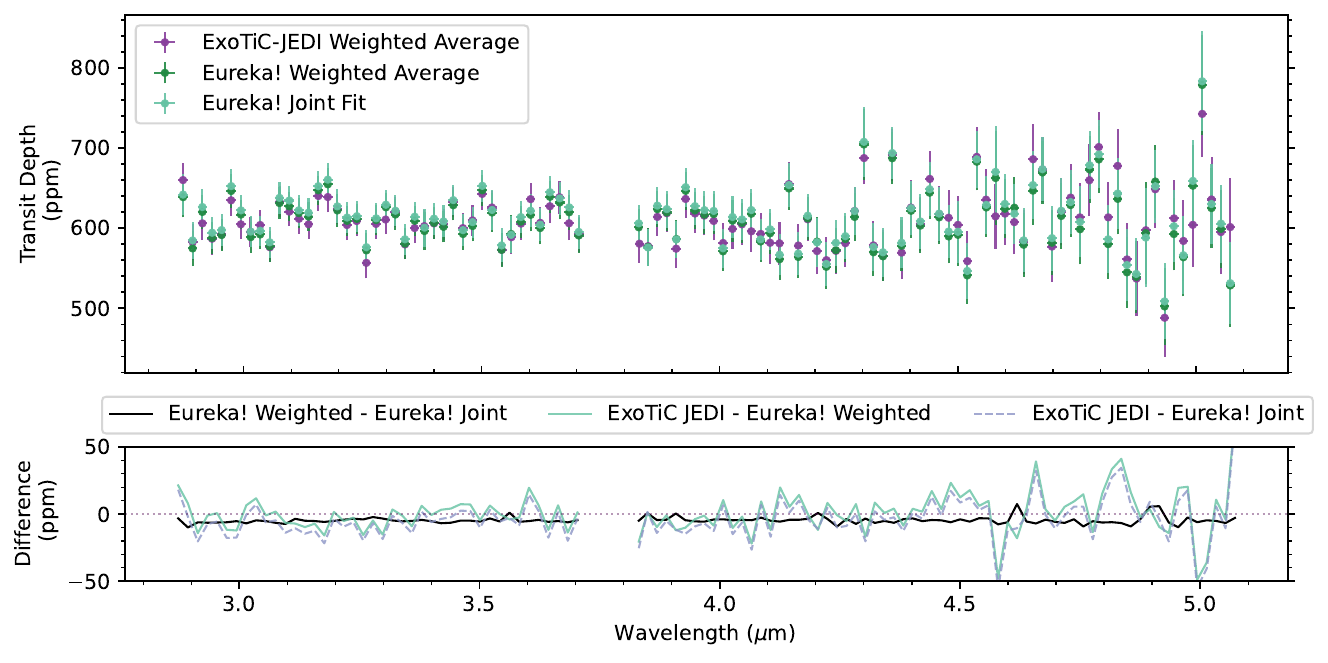}

\caption{Top, Upper Panels: Individual visit transmission spectra for \jedi (left, purples) and \eureka (right, greens). Lower Panel: Difference between individual visit transmission spectra for \jedi (left) and \eureka (right). On average, the \jedi reductions for visit 1 and visit 2 are consistent to within 39\,ppm, while the \eureka reductions agree to within 48\,ppm. The \jedi and \eureka reductions are consistent with each other to within the median transit depth uncertainty for both visit 1 and visit 2. Bottom, Upper Panel: Weighted average transmission spectrum from the two visits from the \jedi (purple) and \eureka (light green) reductions and joint fit transmission spectrum from the \eureka reduction (dark green). Lower Panel: Difference between each of the combined \jedi and \eureka transmission spectra in ppm. As the difference between the two \eureka methods is less the 5\,ppm (black line), the two \jedi -- \eureka lines are difficult to distinguish (coloured lines). On average, the combined visit \eureka and \jedi spectra are consistent to within 10\,ppm.} 
\label{figure:tspec}   
\end{centering}
\end{figure*} 

\begin{figure}
\begin{centering}
\includegraphics[width=0.48\textwidth]{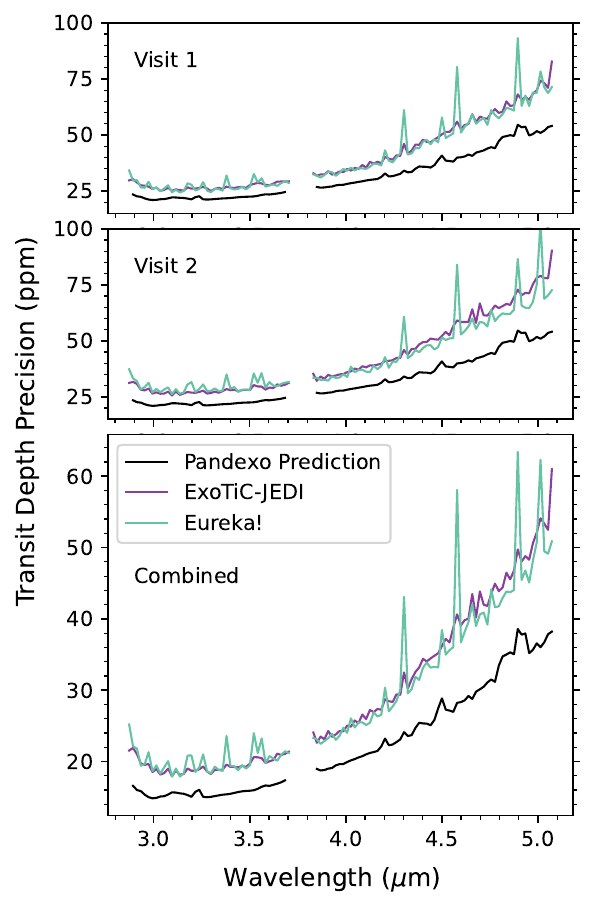}
\caption{Comparison between the transit depth precisions achieved by each reduction and the predicted values from \texttt{PandExo} simulations. We obtain a median transit depth precision in wavelength bins 30 pixels wide ($\sim0.02$\,\micron, $R\sim200$) of 34\,ppm for Visit 1 and 36\,ppm for Visit 2, with both the \jedi weighted and \eureka joint transmission spectra resulting in a median transit depth precision of 25\,ppm. The Eureka! weighted transit depth precisions are indistinguishable from the joint fit precisions, with a median different of less than 0.5\,ppm in each wavelength bin, and therefore are not visible in this plot. In all cases, the average achieved precisions are within 1.3$\times$ the \texttt{PandExo} prediction.} 
\label{figure:pandexo}   
\end{centering}
\end{figure}

The 3--5\,$\mu$m transmission spectra of TOI-836b using a 30-pixel binning scheme for each of the two visits are shown in the upper panels of Figure \ref{figure:tspec}, where no offsets have been applied between NRS1 and NRS2 or between the visits. In general, each visit appears to be consistent between each reduction method, with a median difference in transit depth value of 39\,ppm for \jedi, and 48\,ppm for \eureka, compared to the median transit depth uncertainty for a single visit of 34\,ppm for Visit 1 and 36\,ppm for Visit 2 regardless of reduction method. Each reduction across a single visit are similar, with the median difference between the \jedi and \eureka transmission spectra equal to 11\,ppm for Visit 1 and 17\,ppm for Visit 2, both less than their respective median transit depth precisions. None of the four transmission spectra (two reductions for two visits) shown in the upper panels of Figure \ref{figure:tspec} show any obvious features immediately identifiable by eye as absorption from any expected chemical species in this atmosphere (see \S \ref{section:model_fits}). 
In the lower panels of Figure \ref{figure:tspec}, we also show the weighted average transmission spectrum for both \jedi and \eureka, as well as the joint fit transmission spectrum for \eureka. These combined transmission spectra similarly show no obvious spectral features, and demonstrate consistency between the two reduction pipelines. In particular, the \eureka weighted average and joint transmission spectra have a median transit depth difference of less than 5\,ppm, and produce transit depth precisions within 0.5\,ppm of each other. Given this similarity, we compare the transit depth precisions from \jedi and \eureka for each visit, and in the weighted average case, to the precisions predicted by \texttt{PandExo} \citep{Batalha2017}, shown in Figure \ref{figure:pandexo}. While both reductions are comparable, neither achieves the precision predicted by \texttt{PandExo} across all wavelengths, with the maximum \jedi\, precision 1.6$\times$ and the maximum \eureka precision 2$\times$ the \texttt{PandExo} value. On average, both reductions are within 1.3$\times$ the \texttt{PandExo} value. This is likely due to the more complicated noise presented by the low number of groups required for this extremely bright target (N$_\mathrm{groups}$=3 for TOI-836), as seen in similar NIRSpec programs (e.g., \citetalias{Lustig-Yaeger2023, Moran2023}; \citet{Wallack2024_836.01}).

%%%%%%%%%%%%%%%%%%%%%%%%%%%%%%%%%%%%%%%%%%%%%%%%%%%%%%%%%%%

\section{Interpretation of TOI-836b's Atmosphere} 
\label{section:interpretation}

\begin{figure*}
\begin{centering}
\includegraphics[width=0.9\textwidth]{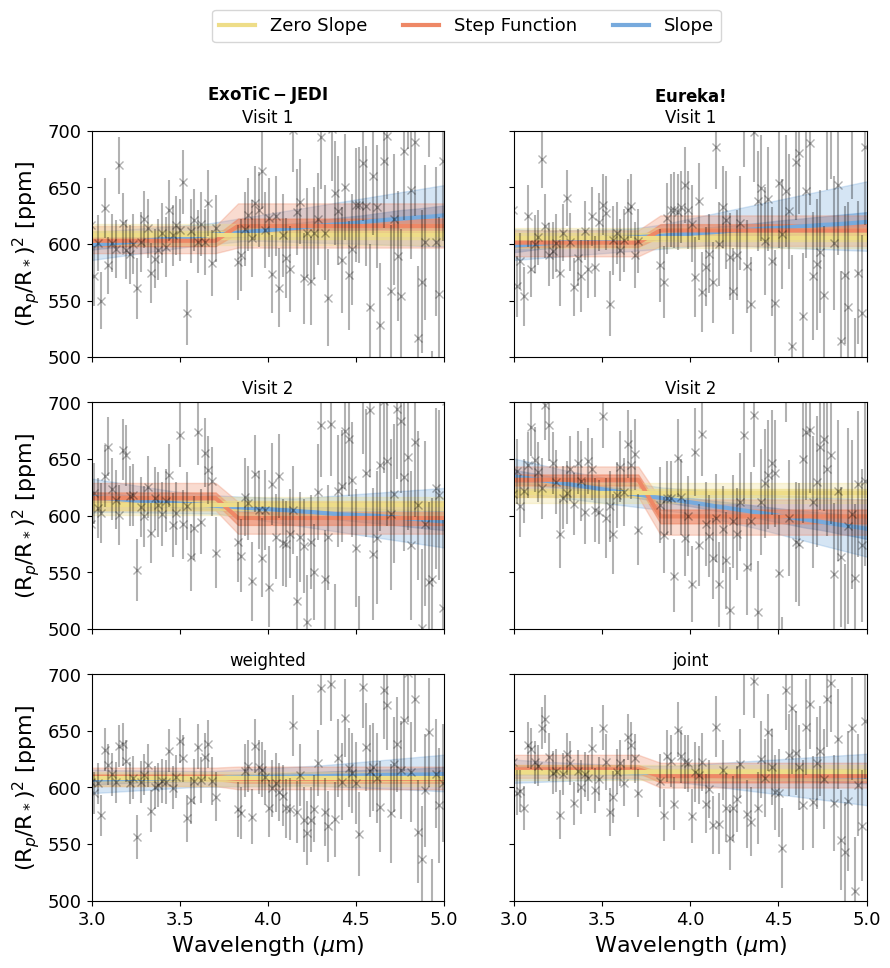}
\caption{Three synthetic fits to the data for both reductions and visits, as well as the combined visit spectra. We use three simple models to fit the data in order to demonstrate the agreement between reductions and visits: 1) a zero-sloped line, 2) a step function to account for offsets between NRS1 and NRS2, and 3) a non-zero-sloped line. Shaded regions illustrate the 1 and 3$\sigma$ bands derived from sampling the posteriors, whereas the line represents the median best-fit model. Overall, the final combined visit spectra are well-fit by a zero-sloped line for both reductions.}  
\label{figure:linefits}   
\end{centering}
\end{figure*} 

To interpret TOI-836b's atmosphere, we must first quantify how well the transmission spectra presented in Figure \ref{figure:tspec} agree with each other across a variety of metrics. In \S \ref{section:synth_fits}, we perform simple synthetic fits to the data to understand: 1) how well the data is fit by a zero (i.e., flat) or non-zero sloped line, 2) potential offsets between NRS1 and NRS2, and 3) whether or not these fits are dependent on the visit and/or reduction method. These choices are driven by structure that has been seen in other small exoplanet atmospheric observations that have complicated the overall interpretation of the atmosphere \citepalias[e.g.,][]{May2023, Moran2023}, and by the fact that the presence of a slope in this wavelength region can be evidence of stellar activity \citep[e.g.,][]{Rackham2018}. Once we are satisfied that we understand whether these concerns may impact our conclusions regarding TOI-836b's atmosphere, we can then move to more physically motivated models. In \S \ref{section:model_fits}, we use \texttt{PICASO} models \citep{Batalha2019, Mukherjee2023} to enable us to understand what region of parameter space we are able to effectively rule out in mean molecular weight and opaque pressure level. 

\subsection{Synthetic Fits to the Data} \label{section:synth_fits}

For the synthetic model fits, we use the \texttt{MLFriends} statistic sampler \citep{MLFriends2016, MLFriends2019} implemented in the open source code \texttt{UltraNest}  \citep{Ultranest}. For each visit and each data reduction, we fit: 1) a one-parameter, zero-slope line, 2) a two-parameter step function composed of two zero-sloped lines, one each for NRS1 and NRS2, and 3) a two-parameter sloped line. The best-fit results are shown in Table \ref{tab:fits} and Figure \ref{figure:linefits}. 

The zero-slope fit for the \jedi Visit 1 and 2 data are consistent to within 1$\sigma$, resulting in a transit depth baseline of $(R_p/R_*)^2$\,=\,608$\pm$3~ppm and 609$\pm$3~ppm for Visit 1 and 2, respectively. The \eureka Visit 1 data is also consistent within the 1$\sigma$ \jedi range at $(R_p/R_*)^2$\,=\,605$\pm$3, however the \eureka Visit 2 data has a somewhat higher ($\sim$2$\sigma$) baseline of $(R_p/R_*)^2\,=\,$620$\pm$4. 

The step function fit for both \jedi and \eureka Visit 1 and 2 are not consistent - Visit 1 has a positive step function, while Visit 2 has a negative step function relative to NRS2. However, comparing across pipelines, the \jedi and \eureka reductions give consistent steps within 1$\sigma$ for each visit. 
For Visit 1 \texttt{ExoTiC-JEDI} and \texttt{Eureka!} produce an offset of +13$\pm$7\,ppm and +11$\pm$7\,ppm, respectively, while for Visit 2, \texttt{ExoTiC-JEDI} and \texttt{Eureka!} produce an offset of -18$\pm$7\,ppm and -31$\pm$7\,ppm, respectively. This leads to a similar discrepancy when fitting the sloped line, where there is agreement across the reduction methods but not from visit to visit. Within 1$\sigma$, both Visit 1 reductions produce a positive slope, while Visit 2 produces a negative slope. This largely suggests that the slope and step function are not astrophysical in nature, unlike those that have been seen in other observations of small exoplanet atmospheres with NIRSpec/G395H \citepalias[e.g.,][]{Moran2023}. Regardless, the size of the offsets obtained for both \jedi visits and for \eureka Visit 1 are smaller than the corresponding median transit depth uncertainty.

To confirm that the apparent offsets between NRS1 and NRS2 need not be a major consideration in our final interpretation of TOI-836b's atmosphere, we can also assess which of the zero-slope, step and slope model is statistically preferred by the data.
Table \ref{tab:fits} lists the likelihoods for each of these fits. For both \jedi and \eureka Visit 1 reductions, and the \jedi Visit 2 reduction, the step function and slope model are not preferred or are only weakly preferred over the zero-slope model - i.e., the data are well described by a flat line given their comparative Bayes factors ($\ln$B$_{12}=\ln$Z$_1$\,[Model 1] $-$ $\ln$Z$_2$\,[Model 2]). For the \eureka reduction of the Visit 2 data, comparisons of the Bayes factors suggest that both the step function and slope model are at least moderately preferred over the zero slope with lnB$_{12}$=6.5 and 2.7\footnote{The rounded integers for \eureka V2 in Table \ref{tab:fits} are -70.45, -63.94, and -67.75 for the zero-slope, step function, and slope models respectively}, respectively. Though Bayes factors do not directly map to $\sigma$-significance for non-nested models \citep{Trotta2008}, for the step function and slope model these roughly translate to a strong and moderate preference, respectively, over the zero slope in this context \citep[see Table 1][]{Trotta2008}. This is an understandable result given that the \eureka Visit 2 offset is larger than the transit depth uncertainty near the detector gap. 

As both \jedi reductions produce consistent results for our synthetic modelling, we proceed with a weighted average transmission spectrum of the two visits for the \jedi reduction. As there is ``moderate'' preference for a step function offset in the \eureka reduction of Visit 2, we choose to leverage the joint fit transmission spectrum from the \eureka reduction for the rest of our analysis.

We also ran our synthetic modelling on the weighted average \jedi and joint fit \eureka transmission spectra to confirm that they are in agreement. In the case of the joint fit, we find that the \eureka data now obtains an offset smaller than the median transit depth uncertainty of 25\,ppm, resulting in the step and slope models no longer being preferred over the zero-slope model. In the case of the weighted average, the \jedi data continues not to prefer the step or slope models, with the calculated offset between NRS1 and NRS2 now consistent with 0\,ppm. The combined visit transmission spectra of TOI-836b is therefore well-described by a flat line regardless of the reduction pipeline used, and we can proceed with our physically motivated modelling.

\begin{table*}[]
\centering
\caption{Results of synthetic fits to Visit 1 \& 2 of both \jedi and \eureka data reductions as well as the combined final transmission spectra. Each column here signifies: for the zero-slope model, the $(R_p/R_*)^2$ baseline intercept in ppm units, 2) for the step function model, the offset between NRS1 and NRS2 in $(R_p/R_*)^2$ ppm units, and 3) for the slope case, the gradient of the slope (ppm/$\mu$m).}
\begin{tabular}{l|ccccc}
                    & \multicolumn{5}{c}{\textbf{Exo-TiC-JEDI (v1/v2/weighted)}}           \\ \hline
\textbf{Model Type} & log Z                & $\chi^2/N$           & Baseline Intercept & NRS1/NRS2 Offset & Slope Gradient\\ \hline
Zero slope          & -62/-66/-68              & 1.10/1.16/1.21            & 608$\pm$3.3/609$\pm$3.5/608$\pm$2.5 & N/A & N/A \\
Step Function       & -64/-65/-72              & 1.07/1.10/1.21            & N/A & +13$\pm$7/-18$\pm$7/-2$\pm$5  &    N/A                  \\
Slope                & -64/-67/-72              & 1.06/1.14/1.21           & N/A & N/A & 13$\pm$6/-10$\pm$6/2$\pm$5  \\ \hline \hline

                    & \multicolumn{5}{c}{\textbf{Eureka! (v1/v2/joint)}}            \\ \hline 
\textbf{Model Type} & log Z                & $\chi^2/N$           & Baseline Intercept & NRS1/NRS2 Offset & Slope Gradient\\ \hline
Zero-slope          & -64/-70/-71              & 1.13/1.25/1.28            & 605$\pm$3.4/620$\pm$4/614$\pm$2.4  & N/A& N/A                         \\
Step Function       & -66/-64/-73              & 1.11/1.07/1.25            & N/A & +11$\pm$7/-31$\pm$7/-6$\pm$5 &  N/A                     \\
Slope                & -67/-68/-75              & 1.10/1.13/1.25            & N/A & N/A & 10$\pm$6/-23$\pm$6/-4$\pm$4  
\end{tabular}
\label{tab:fits}
\end{table*}

\subsection{Ruling out Physical Parameter Space}
\label{section:model_fits}

\begin{figure*}
\begin{centering}
\gridline{\fig{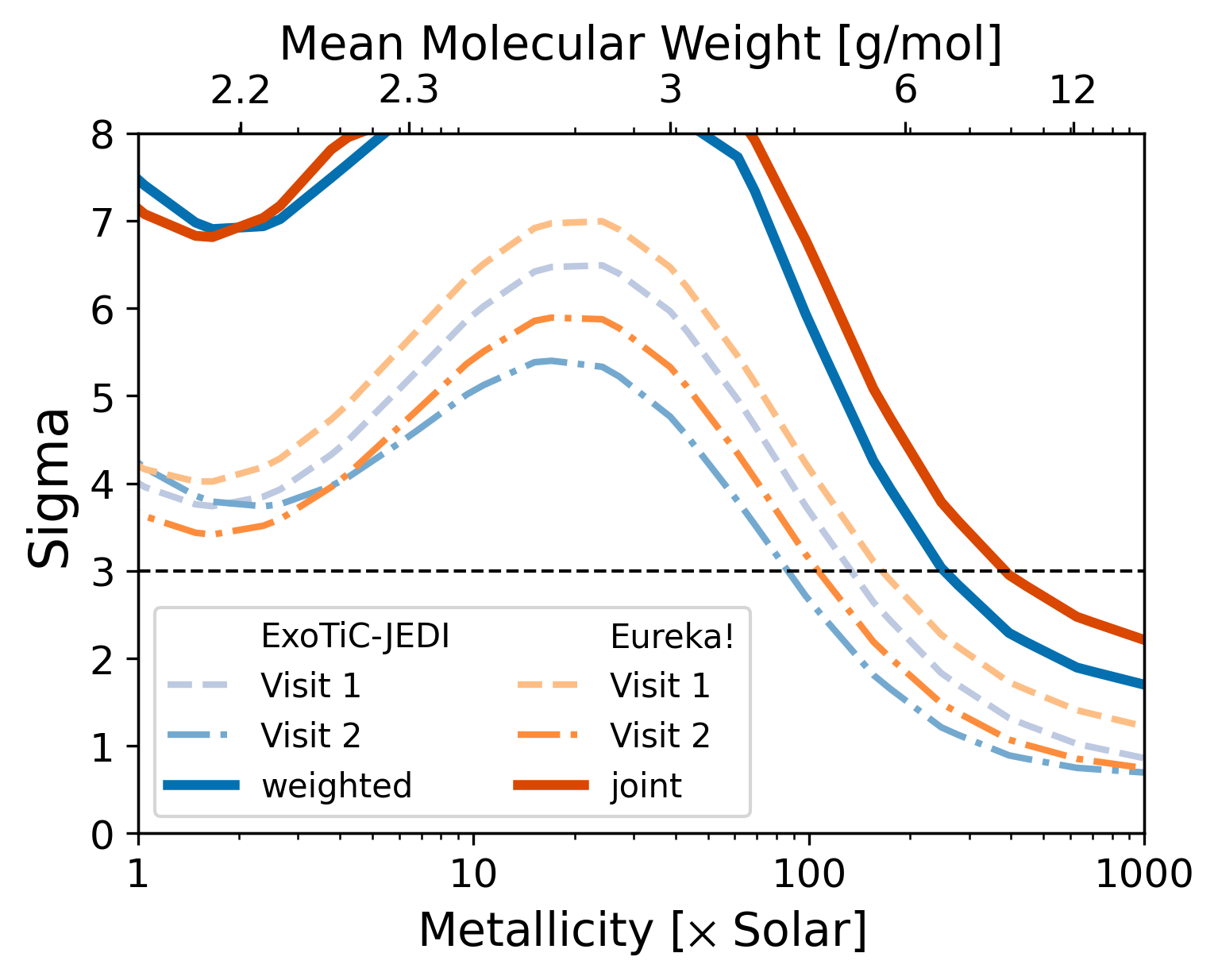}{0.45\textwidth}{(a)}
\fig{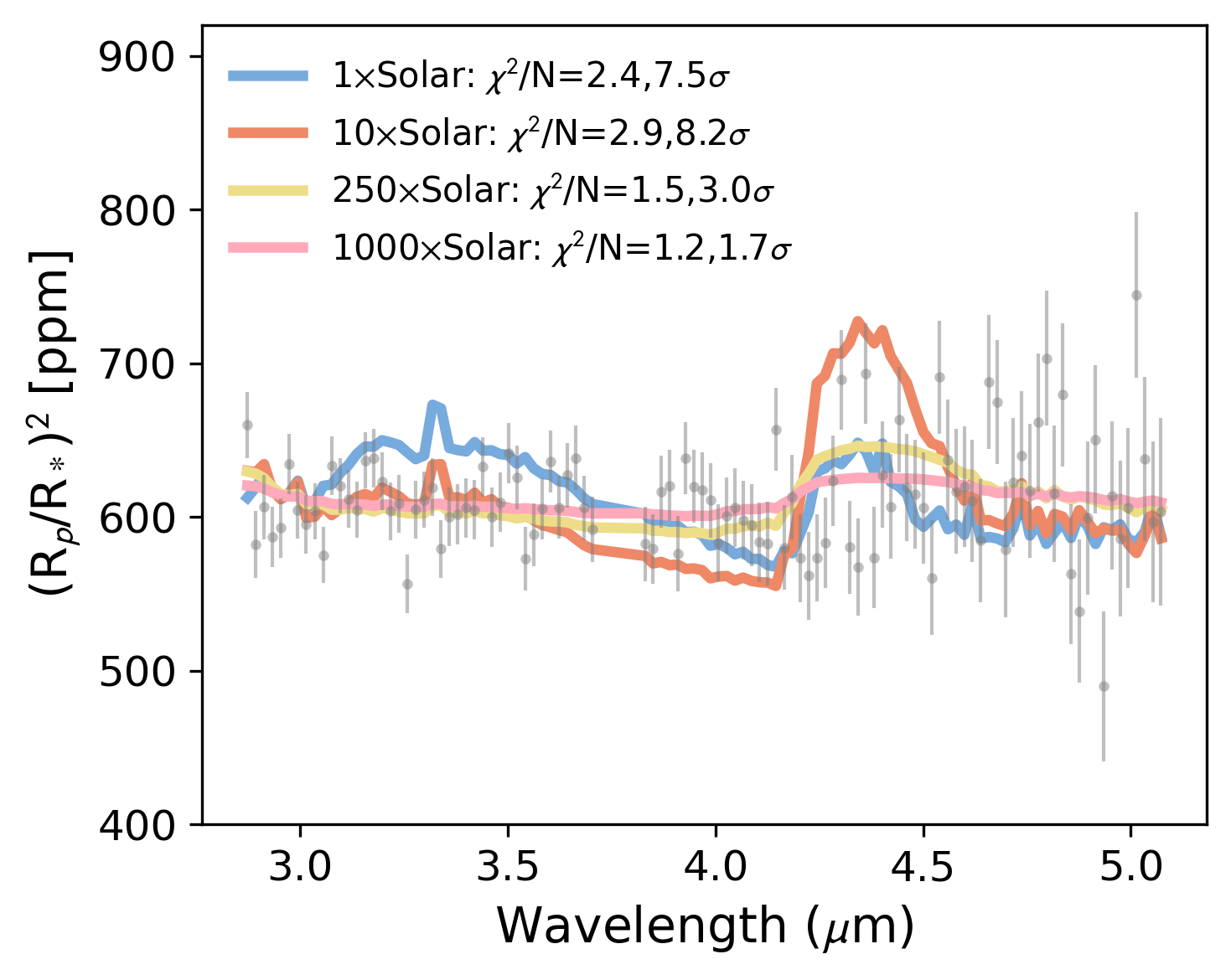}{0.45\textwidth}{(b)}}
\caption{(a) For a single choice in opaque pressure level (0.1 bar) we show the parameter space that can be ruled out in metallicity. Blue lines show the reductions for \jedi (Visit 1, 2, and weighted) and orange lines show the reductions for \eureka (Visit 1, 2, and joint). The black-dashed line indicates the 3$\sigma$ level, below which we are unable to confidently rule out models. Ultimately our data rules out metallicities $<250\times$Solar, corresponding to a mean molecular weight of $\sim6$\,g\,mol$^{-1}$. (b) For four of the metallicity cases shown in (a), we show the spectra relative to the weighted data from \jedi. We also indicate the $\chi^2$/N and $\sigma$ for reference. } 
\label{figure:mhcld}   
\end{centering}
\end{figure*}

The spectral feature sizes of transmission spectra are largely driven by the scale height (=kT/$\mu$g), and potential muting by aerosols \citep[e.g.][]{Sing2016}. In order to understand what region of parameter space we can rule out for this system, we create a grid of spectral models as a function of metallicity and ``opaque pressure level''. For the former parameter, metallicity, it is unlikely that this system (R\,=\,1.7~$R_\oplus$) has a large hydrogen-helium envelope. However, similar to the analysis of the TRAPPIST-1 system \citep{Moran2018} and of other small planets \citepalias[e.g.][]{Moran2023,Lustig-Yaeger2023}, metallicity offers a suitable proxy for the mean molecular weight. For example, for our given assumption in temperature-pressure profile, 100$\times$Solar corresponds to a mean molecular weight of 4.3\,g\,mol$^{-1}$, while 1000$\times$Solar corresponds to a mean molecular weight of 15.7\,g\,mol$^{-1}$. 
The other parameter, ``opaque pressure level'', is a term adapted from \citetalias{Lustig-Yaeger2023} and represents a pressure below which the atmosphere is opaque \citep[e.g.,][]{Seager2000, Charbonneau2002, Berta2012, Kreidberg2014, Knutson2014}\footnote{\citetalias{Lustig-Yaeger2023} used the term ``apparent surface pressure'' but it has the same meaning}. Other manuscripts have referred to this as a ``cloud top pressure'' \cite[e.g.,][]{Kreidberg2014,Moran2018}. Here we use a more general term as we cannot differentiate between a cloud top pressure from a surface pressure.

Using a range of metallicities (1--1000$\times$Solar, log spaced with 26 grid points) and opaque pressure levels (100--10$^{-4}$ bar, log-spaced with 5 grid points), we compute a grid of transmission spectra using the open source \texttt{PICASO} package \citep{Batalha2019}. For the pressure-temperature profile, we use a 1D 5-parameter double-grey analytic formula \citep{Guillot2010}. We also test whether or not the results are sensitive to our choice of pressure-temperature profile by computing a similar grid with simple isothermal pressure-temperature profiles. Our conclusions do not change depending on this choice. Given the pressure-temperature profile, we fix the elemental ratio C/O to solar (=0.55 \citealt{Asplund2009}) and obtain the chemistry by interpolating on a pre-computed chemical equilibrium grid. The chemistry grid was computed by \citet{Line2013} with NASA's CEA code \citep{gordon1994computer}. This grid is publicly available on \texttt{GitHub} as part of \texttt{CHIMERA}'s open source code\footnote{\url{https://github.com/mrline/CHIMERA}}. Of note to this analysis, the molecules which absorb from 3--5\,$\mu$m are H$_2$O, CH$_4$, CO$_2$, and CO, which are all included in the grid. 

Figure \ref{figure:mhcld} shows the results of the grid analysis, where we show how our confidence level, expressed as a $\sigma$-level, changes as a function of metallicity for an intermediate opaque pressure level of 0.1 bar. The general behaviour of the significance curves in Figure \ref{figure:mhcld}a, which peak toward 10-50$\times$Solar and decrease toward 1000$\times$Solar, is well-documented \cite[e.g.][]{Moran2018}. Toward 10$\times$Solar the magnitude of spectral features increases because the added molecular opacity is better able to surpass the contribution from H$_2$/He continuum without affecting the mean molecular weight, increasing the significance to which the spectral features can be ruled out. Beyond $\sim$50$\times$Solar the contribution from the heavier metals starts to increase the mean molecular weight of the atmosphere and results in overall smaller spectral features which are harder to more confidently rule out. 

We choose to show 0.1 bar for reference as it is synonymous with the tropopause of all Solar System objects \citep{Robinson2014}. Here, $\sigma$ is computed by converting $\chi^2$/N to a p-value, and then to a $\sigma$-significance. For each individual visit, we are able to rule out metallicities lower than 100--160$\times$Solar depending on the visit and the reduction. For example, in the case of the first visit, the individual \jedi and \eureka spectra enable metallicities to be ruled out at the $<$130$\times$Solar and $<$160$\times$Solar-level, respectively. With the combined spectra, we are further able to rule out metallicities $<250\times$Solar and $<380\times$Solar, for \jedi and \eureka, respectively, corresponding to mean molecular weights of $\sim$6--9\,g\,mol$^{-1}$. Combining both visits enables us to rule out nearly double the parameter space in metallicity, demonstrating the potential of multi-visit observing strategies. As shown in Figure \ref{figure:mhcld}a, the reported 3$\sigma$ lower limit on metallicity is dependent on the reduction method. However, the overall scientific conclusions are agnostic to the data reduction, as in all cases we are able to rule out H$_2$-dominated atmospheres with mean molecular weights less than $\sim$6\,g\,mol$^{-1}$. 

Figure \ref{figure:mhcld} shows the results for an intermediate cloud case, though we ran a full grid of both cloud-free and highly cloudy cases. For an effective cloud-free atmosphere, in which the atmosphere only becomes opaque below 100 bar, the combined \jedi spectrum is able to rule out the 300$\times$Solar case. For cases where the opaque pressure level is 10$^{-4}$ bar, comparable to a highly lofted cloud, we can rule out cases $\le$100$\times$Solar. Both the 100 bar and 10$^{-4}$ bar cases represent unlikely physical scenarios, as clouds are expected to form in super-Earth atmospheres \citep{Mbarek2016} (i.e., atmospheres are highly unlikely to be effectively cloud free), and clouds lofted to high altitudes are unlikely to be completely opaque \citep{Robinson2014}, however as end-member cases they demonstrate that solar-like composition atmospheres are not plausible for TOI-836b regardless of the height of any opaque pressure level. Additionally, we tested whether or not our conclusions are affected by the choice of binning scheme and found that the conclusions are unchanged. 

Figure \ref{figure:mhcld}b shows four of the spectra used to compute the $\sigma$-significance curves in \ref{figure:mhcld}a, for reference, along with the weighted average spectra from \jedi. The main features shown are that of CH$_4$ and CO$_2$, in the 1$\times$Solar case, and primarily H$_2$O and CO$_2$ in the other cases. Figure \ref{figure:mhcld}b also lists the $\chi^{2}$/N and sigma rejection thresholds for each of the metallicity cases, demonstrating, for example, that we cannot confidently distinguish between atmospheres with 250$\times$Solar and 1000$\times$Solar metallicities given the constraints we achieve with two combined transit observations.

%%%%%%%%%%%%%%%%%%%%%%%%%%%%%%%%%%%%%%%%%%%%%%%%%%%%%%%%%%%

\subsection{Theoretical predictions of possible interiors of TOI-836b}

\begin{figure}
\begin{centering}
\includegraphics[width=\linewidth]{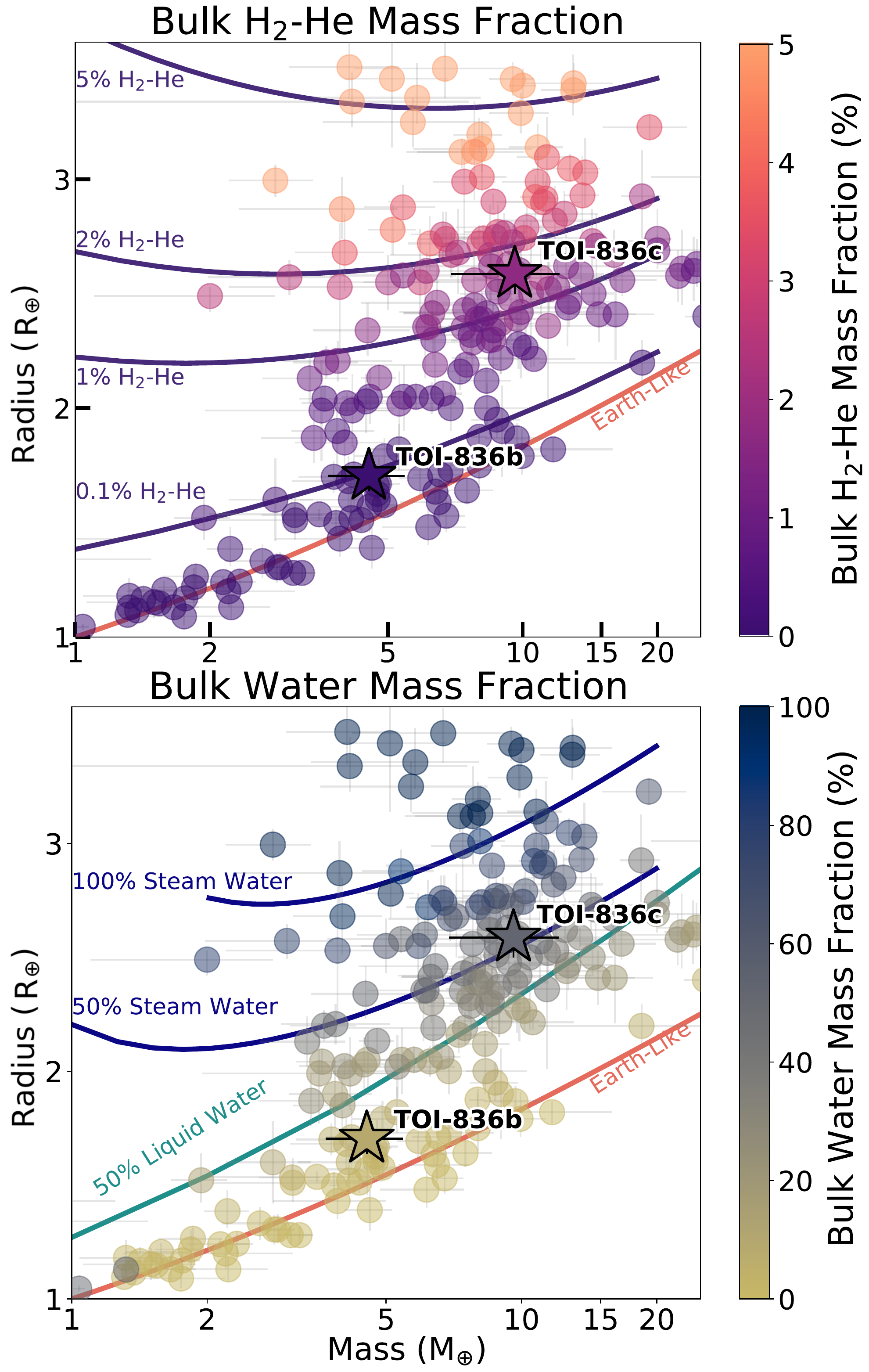}
\caption{Mass-radius plots for the population of small exoplanets demonstrating possible interior compositions as calculated by \texttt{SMINT}, where the colour of each marker represents either the bulk H$_2$-He (top) or bulk water (bottom) mass fractions. The planets in the TOI-836 system are denoted by star-shaped markers. Density curves for the Earth-like and 50 wt\% liquid water compositions \citep{Zeng2016}; 50\% and 100\% steam atmospheres assuming $T_{\mathrm{eq}}=600$ K and an Earth-like core \citep{Aguichine2021}; and 0.1\%, 1\%, 2\% and 5\% H$_2$-He composition assuming an age of 5 Gyr \citep{Lopez2014} are also plotted for reference. The background planet population is obtained from the NASA Exoplanet Archive.}
\label{figure:interiors}   
\end{centering}
\end{figure}

TOI-836b's mass and radius place it at a very intriguing position in the mass-radius diagram towards the lower edge of the radius valley. Furthermore, despite its low density, its parameters are compatible with a pure rock composition (no iron core) at the $1\sigma$ level. To infer the bulk properties of TOI-836b, we use the \texttt{SMINT} (Structure Model INTerpolator) package from \cite{Piaulet2021}, which performs an MCMC retrieval of planetary bulk compositions from pre-computed grids of theoretical interior structure models. We consider two possible compositions for the interior: 1) an Earth-like core with a H\textsubscript{2}-He envelope of solar metallicity \citep{Lopez2014}, and 2) a refractory core with a variable core mass fraction and a pure H$_2$O envelope and atmosphere on top \citep{Aguichine2021}. These compositions represent end-member cases between an envelope that would form by accreting nebular gas with a Sun-like composition, and a high mean molecular weight envelope where water is used as a proxy for all volatiles. Based on its bulk properties and these scenarios, we find that TOI-836b could have an envelope mass fraction of at most $0.1\%$ in the case of gas of solar composition, or a water mass fraction of $9\pm 5\%$ in the pure H$_2$O case as shown in Figure \ref{figure:interiors}.

%%%%%%%%%%%%%%%%%%%%%%%%%%%%%%%%%%%%%%%%%%%%%%%%%%%%%%%%%%%

\section{Discussion} 
\label{section:discussion}
%%%%%%%%%%%%%%%%%%%%%%%%%%%%%%%%%%%%%%%%%%%%%%%%%%%%%%%%%%%

\subsection{The TOI-836 System}

The two planets of the TOI-836 system are located on opposite sides of the radius valley, which translates into very different bulk compositions when inferred from interior structure models. The interior of TOI-836b is compatible with a pure rock (no iron core) composition at $1\sigma$, meaning that the possibility that TOI-836b is a terrestrial planet cannot be excluded. From our interior modelling, we also find that the planet could be made of at most $0.1\%$ solar metallicity gas or $9\pm 5\%$ pure water. In contrast, the possible bulk compositions of TOI-836c are $1.74^{+0.55}_{-0.48} \%$ in the case of solar metallicity gas, or $52^{+15}_{-14} \%$ in the pure water case \citep{Wallack2024_836.01}. Figure \ref{figure:interiors} shows these possible interior compositions for both planets, which represent end-member cases for hydrogen-dominated and pure water atmospheres such that intermediate compositions are also possible.

Given TOI-836b's high equilibrium temperature and the stellar insolation flux received, photoevaporation could likely be responsible for the observed lack of a hydrogen-dominated atmosphere. Applying the photoevaporation model of \citet{Rogers2021} to a planet with the properties of TOI-836b, and including a core mass of $4.5~M_{\oplus}$, we find that any initial envelope mass fraction in the range 2--30\% is blown away in $<$400 Myr. Applying the model to a planet with the properties of TOI-836c, and including a core mass of $9.4~M_{\oplus}$, we find that hydrogen envelopes of up to 10\% can be retained. Given the reported age of TOI-836 of 5.4$^{+6.3}_{-5.0}$\,Gyr \citep{Hawthorn2023}, this analysis strongly suggests the absence of a hydrogen-dominated atmosphere for TOI-836b, which is in line with the apparent $>$6\,g\,mol$^{-1}$ mean molecular weight derived in \S \ref{section:model_fits}. The bulk composition of TOI-836c, however, is still degenerate, with low mean molecular weight atmospheres still plausible, particularly in the presence of clouds and hazes \citep{Wallack2024_836.01}. 
These models provide the first clues as to this system's possible evolution, although we caution that transmission spectra alone are unable to determine whether the TOI-836 planets formed with their current masses, or if the present-day difference in bulk compositions is the consequence of photoevaporation.

\subsection{Implications for Future JWST Observations}
\begin{figure}
\begin{centering}
\includegraphics[width=0.45\textwidth]{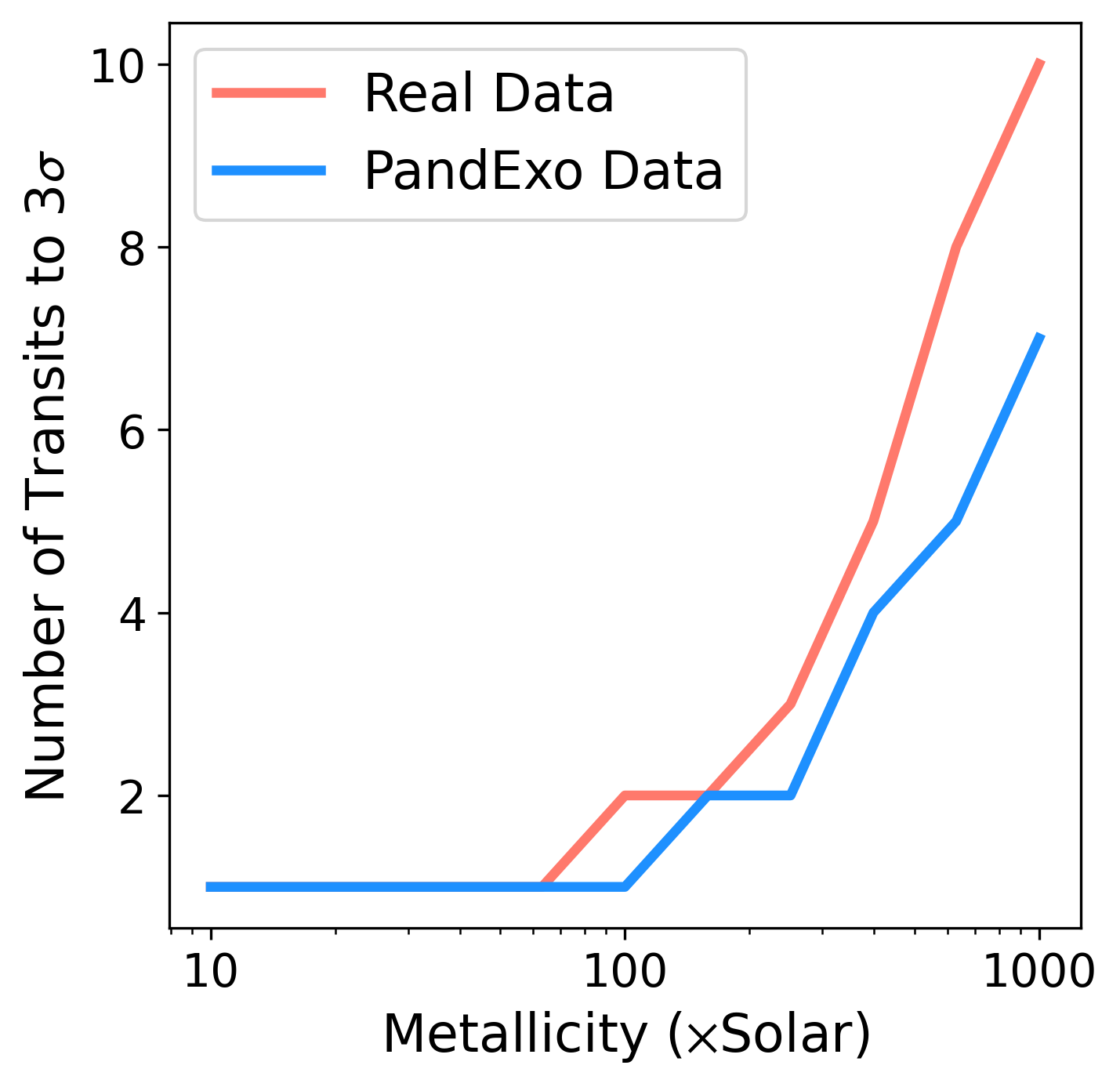}
\caption{The number of transits needed to rule out a zero-sloped line at 3$\sigma$ as a function of metallicity for an opaque pressure level of 0.1 bar. Here, ``real data'' uses the precision derived from the first visit of the \texttt{Eureka!} reduction and assumes that additional transits improve the precision based on the measured improvement from a single visit to the \texttt{Eureka!} joint fit. The \texttt{PandExo} data curve follows simulations computed with the JWST simulation tool \texttt{PandExo}. } 
\label{figure:moreobs}   
\end{centering}
\end{figure}

Here we examine how our results can inform the planning of future observations of small planets with high atmospheric metallicities that orbit bright stars, comparing our measured data to \texttt{PandExo} JWST simulations, which are used by the community for planning observations. Using the grid of models described in \S \ref{section:model_fits} we determine how many additional transits would be needed to rule out a certain metallicity model. This is an identical exercise to that performed in Figure \ref{figure:mhcld}, except here our ``data'' is a \texttt{PandExo} simulation of a featureless spectrum. We compute noise simulations with \texttt{PandExo} \citep{Batalha2017} using an identical observational setup to our program here, assuming that each additional visit provides a gain in precision of $\sqrt{\mathrm{n}\_\mathrm{transit}}$. For the ``real data'' case, we use the noise budget measured in this program from Visit 1,  assuming that the increase in precision is equivalent to what we have measured moving from individual visits to the combined joint \eureka reduction. Doing so results in a measured precision gain of $\sim$98\%$\sqrt{2}$ for TOI-836b. Using each of the estimates for noise, we compute simulated observations of each modelled spectrum and then compute the number of transits needed to rule out a zero-sloped line with 3$\sigma$ confidence. Additionally, we include random noise and repeat the test for 1000 different random noise instances. Figure \ref{figure:moreobs} shows the median result for the case of an opaque pressure level of 0.1 bar. 

Overall, ruling out cases up to 1000$\times$Solar metallicity for TOI-836b would require up to eight additional transits, assuming that the data continued to result in a precision gain of $\sim$98\%$\sqrt{n\_\mathrm{transit}}$. For lower metallicity cases ($<$100$\times$Solar), \texttt{PandExo} predictions are in line with those based on the real data. However, for higher metallicities, \texttt{PandExo} data appears to be somewhat optimistic, resulting in estimates requiring 1-2 fewer transits than the real data. This result should be noted for future observation planning of planets that are expected to be heavily enriched in metals, particularly for those around bright stars with more complex noise properties. 

\section{Conclusions}
\label{section:conclusions}

We have presented two JWST NIRSpec/G395H observations of the transmission spectrum of the super-Earth TOI-836b. We produce two reductions of the data with independent pipelines, \jedi and \eureka, resulting in a median transit depth uncertainty for both methods of 34\,ppm for Visit 1 and 36\,ppm for Visit 2 in 30 pixel wide bins. We combine our two visits using a weighted average for \jedi and a joint fit for \eureka, and find a combined median transit depth precision of 25\,ppm in both cases. We also find sub-ppm differences in the precision obtained by the \eureka joint fit and a weighted average of the individual \eureka visits at all wavelengths. 

When modelling our transmission spectra, we find that transmission spectra that appear to be flat ``by-eye'' can have different retrieved transit depth baselines and detector offsets. We caution that these model parameterisations are a simple and basic test to determine first-order structures in the data, but are not necessarily an accurate representation of the intrinsic scatter across the transmission spectrum. Future work will be needed as more data is collected to better characterise the noise properties being seen in JWST observations. Careful analyses of each visit and each data reduction method should therefore be done individually and assessed collectively, even if the data appear to be overall consistent to within 1$\sigma$.

Our final combined transmission spectrum from each reduction method is well described by a flat line, with no obvious atmospheric features. \texttt{PICASO} modelling enables us to rule out atmospheres of at least $<$100$\times$Solar metallicity regardless of the height of an opaque pressure level (equivalent to either a cloud deck or surface). With our combined two visit spectra for the 0.1 bar case, we specifically rule out $<$250$\times$Solar metallicities for the \jedi spectrum and $<$380$\times$Solar metallicities for the \eureka spectrum. These constraints allow us to rule out atmospheres with mean molecular weights less than $\sim6$\,g\,mol$^{-1}$. Given the modelling setup considered in this work, combining both visits enables us to rule out nearly double the parameter space in metallicity when compared to each visit individually. Comparing our mean molecular weight for TOI-836b to interior and photoevaporation evolution modelling strongly supports our overall conclusion that this super-Earth does not possess a H$_2$-dominated atmosphere, in possible contrast to the larger, exterior TOI-836c. 

As JWST continues to observe small planets, we recommend that care is taken when using simulation tools to determine how many transits may be needed to rule out certain physical scenarios, particularly in the case of observations that require a small number of groups. For high-metallicity atmospheres ($>$100$\times$Solar), we find that \texttt{PandExo} predictions are optimistic compared to the precision gains from our measured data, and yield estimates with 1–2 transits less than may be required. This should be accounted for in future observation proposals of small planets with JWST. \linebreak

%%%%%%%%%%%%%%%%%%%%%%%%%%%%%%%%%%%%%%%%%%%%%%%%%%%%%%%%%%%

\noindent We thank the anonymous referee for their comments that helped improve the quality and clarity of this paper.
The data products for this manuscript can be found at the following Zenodo repository: 10.5281/zenodo.10658637.
L.A. would like to thank D. Grant and M. Radica for useful discussions regarding JWST data analysis.
This work is based on observations made with the NASA/ESA/CSA James Webb Space Telescope. The data were obtained from the Mikulski Archive for Space Telescopes at the Space Telescope Science Institute, which is operated by the Association of Universities for Research in Astronomy, Inc., under NASA contract NAS 5-03127 for JWST. These observations are associated with program \#2512. Support for program \#2512 was provided by NASA through a grant from the Space Telescope Science Institute, which is operated by the Association of Universities for Research in Astronomy, Inc., under NASA contract NAS 5-03127.
L.A. acknowledges funding from STFC grant ST/W507337/1 and from the University of Bristol School of Physics PhD Scholarship Fund.
This work is funded in part by the Alfred P. Sloan Foundation under grant G202114194.
Support for this work was provided by NASA through grant 80NSSC19K0290 to J.T. and N.W. 
H.R.W. was funded by UK Research and Innovation (UKRI) under the UK government’s Horizon Europe funding guarantee [grant number EP/Y006313/1].
This material is based upon work supported by NASA’S Interdisciplinary Consortia for Astrobiology Research (NNH19ZDA001N-ICAR) under award number 19-ICAR19\_2-0041.
This work benefited from the 2023 Exoplanet Summer Program in the Other Worlds Laboratory (OWL) at the University of California, Santa Cruz, a program funded by the Heising-Simons Foundation. 
This research also made use of the NASA Exoplanet Archive, which is operated by the California Institute of Technology, under contract with the National Aeronautics and Space Administration under the Exoplanet Exploration Program.

Co-Author contributions are as follows: 
LA led the data analysis and write-up of this study. NEB led the modelling efforts. NLW and JIAR provided reductions and analyses of the data. AA performed the interior modelling. 
HRW advised throughout the analysis and manuscript preparation. All authors read and provided comments and conversations that greatly improved the quality of the manuscript.

\software{\texttt{astropy} \citep{Astropy2013, astropy, AstropyCollaboration2022}, \texttt{batman} \citep{Kreidberg2015}, \texttt{emcee} \citep{Foreman-Mackey2013}, \texttt{Eureka!} \citep{Bell2022},  \texttt{ExoTiC-Jedi} \citep{Alderson2022}, \texttt{ExoTiC-LD} \citep{Grant2022}, \texttt{Matplotlib} \citep{matplotlib},  \texttt{NumPy} \citep{numpy}, \texttt{pandas} \citep{pandas}, \texttt{PICASO} \citep{Batalha2018, Mukherjee2023}, \texttt{PandExo} \citep{Batalha2017},  
\texttt{scipy} \citep{scipy}, \texttt{SMINT} \citep{Piaulet2021}, STScI JWST Calibration Pipeline \citep{Bushouse2022}, \texttt{ultranest} \citep{Ultranest}, \texttt{xarray} \citep{xarray}} \newline

\facilities{JWST (NIRSpec)} 

The JWST data presented in this paper were obtained from the Mikulski Archive for Space Telescopes (MAST) at the Space Telescope Science Institute. The specific observations analysed can be accessed via DOI: 10.17909/fpwt-rn60.

\bibliography{references}{}
\bibliographystyle{aasjournal}

\end{document}